\documentclass[letterpaper]{jpconf}
\usepackage{graphicx}
\newcommand{\met}{\mbox{$E_{T}\!\!\!\!\!\!\!/\,\,\,\,$}}
\begin{document}
\title{Search for New Physics at CDF}

\author{John Strologas}

\address{University of New Mexico \\(on behalf of the CDF Collaboration)}

\begin{abstract}
We present the current status of the search for new physics at CDF, using integrated luminosity up to 3.2 fb$^{-1}$.  We cover searches for supersymmetry, extra dimensions, new heavy bosons, and generic dilepton resonances.

\end{abstract}
\section{Introduction}

Although the Standard Model (SM) is extremely successful in describing the known particles and their interactions, it is not the final theory.  Many open questions are not explained by the current framework.  The Planck and the electroweak scales are 17 orders of magnitude apart, a fact that makes the theory unstable unless it is fine-tuned.  Gravity has not been incorporated yet in a consistent way.  The dark matter and dark energy are not explained.  There are several asymmetries that we cannot account for, including the boson-fermion and the particle-antiparticle asymmetries. In addition, we cannot understand the necessity for more-than-one particle interactions and we would like to see them unified at a higher energy-scale.  Either an extension of the SM or a radically new approach is needed in order to address these known problems.

At CDF, a diverse new-physics program investigates possible signal predicted by new theories and also considers model-independent experimental signatures inconsistent with the SM.  In this paper we present the latest searches for supersymmetry, extra dimensions, and new heavy bosons.  Additional analyses, including signature-based ones, can be found at the public web pages of the experiment [1].
%\vspace{-0.1cm}
\section{Supersymmetry}

Supersymmetry (SUSY) predicts a fermion for every known boson and vice-versa.  The superpartners differ only in their spin.  This symmetry is obviously broken, resulting in much heavier SUSY particles.  The two leading SUSY breaking scenarios are mSUGRA (SUSY-breaking is communicated through gravity; the lightest SUSY particle (LSP) is the neutralino) and GMSB (SUSY-breaking is communicated with gauge fields; the LSP is the gravitino).  If $R$-parity is conserved the SUSY particles will be produced in pairs and the LSP will be stable, associating SUSY events with large missing transverse energy ($\met$).  CDF has performed searches for charginos ($\tilde{\chi}^{\pm}_{1}$), neutralinos ($\tilde{\chi}^{0}_{1,2}$), squarks ($\tilde{q}$), gluinos ($\tilde{g}$) and sneutrinos ($\tilde{\nu}$), interpreted in the mSUGRA scenario with $R$-parity conservation, searches for neutralinos in the GMSB scnenario and searches for $R$-parity violating (RPV) $\tilde{\nu}$ production.

\subsection{Search for Charginos and Neutralinos}

The $\tilde{\chi}^{\pm}_{1}\tilde{\chi}^{0}_{2}$ associated production cross section of the order of 0.1-1 pb has not been excluded yet.  At the same time, the trilepton decay signature is characterized by very low SM backgrounds, making the trileptons+$\met$ the golden channel for the discovery of SUSY at the Tevatron.

\begin{figure}[t]
\begin{minipage}{18pc}
\includegraphics[scale=0.42]{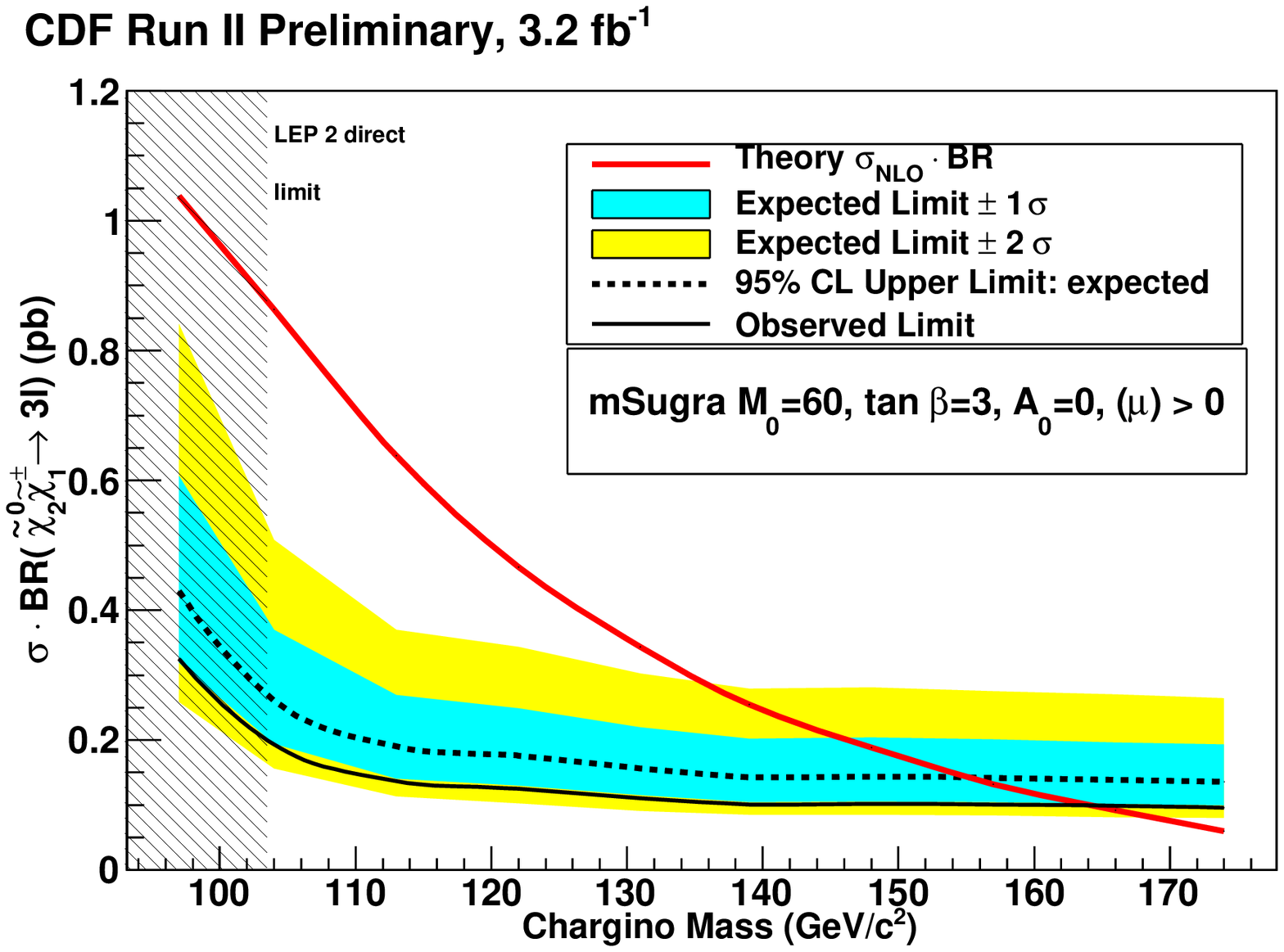}
\caption{The mSUGRA $\sigma\cdot\rm{BR}$ vs. chargino mass limit from the trilepton analysis.\label{fig1}}
\end{minipage}\hspace{1pc}%
\begin{minipage}{18pc}
\includegraphics[scale=0.40]{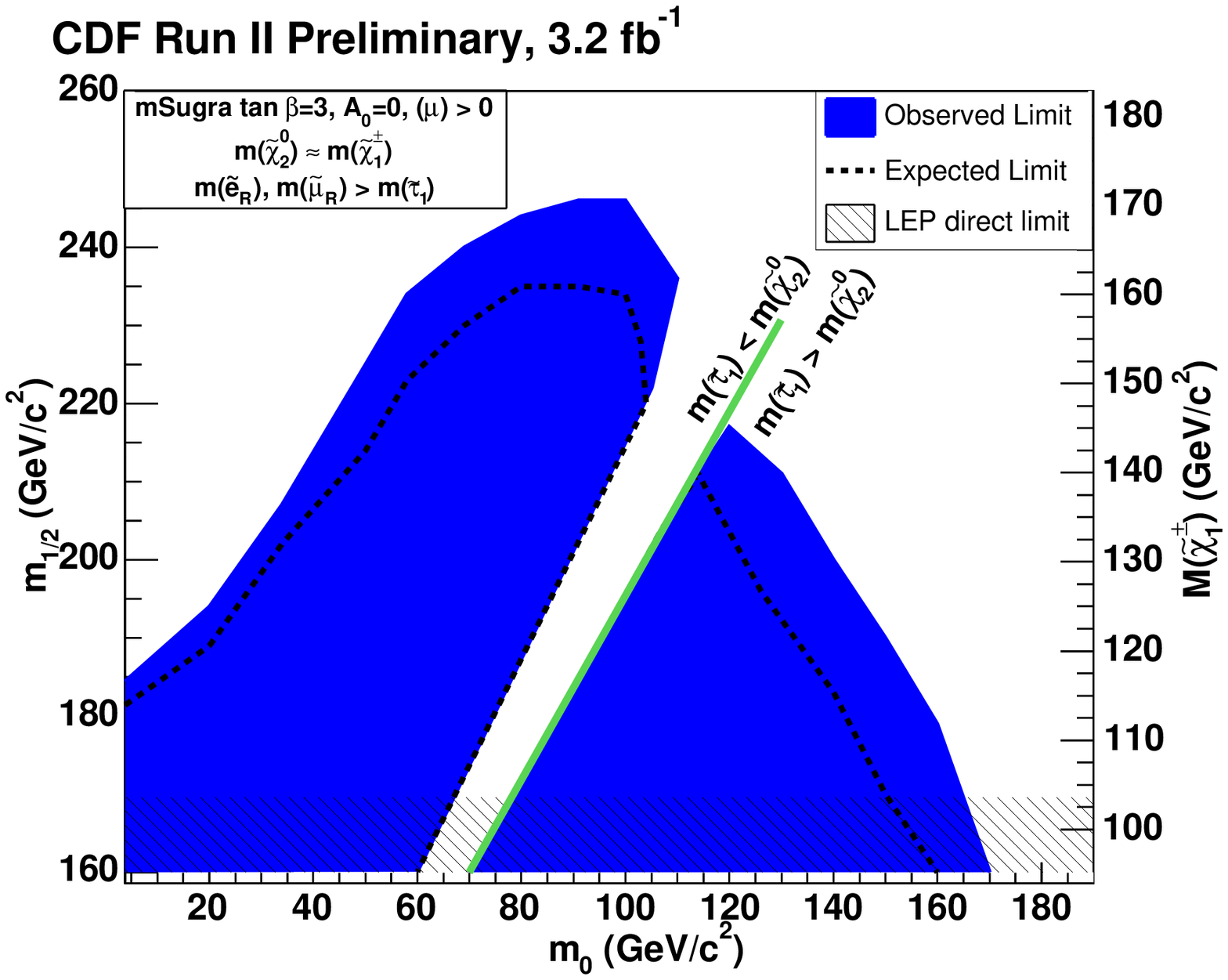}
\caption{The mSUGRA $m_0$ vs. $m_{1/2}$ limit from the trilepton analysis. \label{fig2}}
\end{minipage}
\label{wcharge}
\end{figure}

We analyze 3.2 fb$^{-1}$ of CDF data by selecting either three leptons or two leptons accompanied by an isolated track (which can be the result of a soft lepton or a single-prong tau decay).  We require $\met>20$ GeV (for Drell-Yan (DY) and QCD rejection), $N_{\rm jets} \le 1$ and $H_T<80$ GeV -- where $H_T$ is the scalar sum of the transverse energies of all the jets in the event -- (for top-antitop rejection), dilepton mass above 20 GeV/$c^2$ (for heavy-flavor and resonance rejection and for LSP acceptance) and outside the Z mass window (for DY and diboson reduction).  The main trilepton backgrounds are diboson (61\%), $Z+\gamma$ (22\%) , DY+fake (15\%).  We expect $1.5 \pm 0.2$ trilepton and  $9.4 \pm 1.4$ dilepton+track events from the SM.  From a benchmark mSUGRA point ($m_0=60$ GeV/$c^2$, $m_{1/2}=190$ GeV/$c^2$, $\tan\beta=3$, $A_0=0$ and $\mu>0$) we expect $7.4 \pm 0.7$ and $11 \pm 1$ respectively.  We observe 1 and 6 events respectively, in consistence with the SM.  For $m_0=60$ GeV, $A_0=0$, $\tan\beta=3$ and $\mu>0$, we can set the limit of $M_{\tilde{\chi}_1^{\pm}}>164$ GeV/$c^2$ at 95\% confidence level (CL), as seen in Figure \ref{fig1} (expected limit is 155 GeV/$c^2$). If we fix only the $\tan\beta$, $A_0$ and $\mu$ parameters at the above values, we can exclude a region in the 2-dimensional $m_0$ vs. $m_{1/2}$ space, as seen in Figure \ref{fig2}.

\subsection{Search for Squarks and Gluinos}

The dominant $\tilde{q}-\tilde{g}$ production process depends on their masses.  If $M_{\tilde{q}} \ll M_{\tilde{g}}$, we expect mainly di-squark production, if $M_{\tilde{q}} \gg M_{\tilde{g}}$, we expect mainly di-gluino production and if $M_{\tilde{q}} \approx M_{\tilde{g}}$, we expect in addition squark-gluino production.  The experimental signature is 2-jets+$\met$, 4-jets+$\met$ and 3-jets+$\met$ respectively.  Although the production is strong (QCD) it suffers by limited available kinematic phase space and large QCD multijet and ($W/Z$)+jets backgrounds.  For this reason, the analysis is broken-down in jet-multiplicity bins and optimized for each bin separately.  The optimization variables are $\met$ and $H_T$.  

\begin{figure}[t]
\begin{minipage}{18pc}
\includegraphics[scale=0.37]{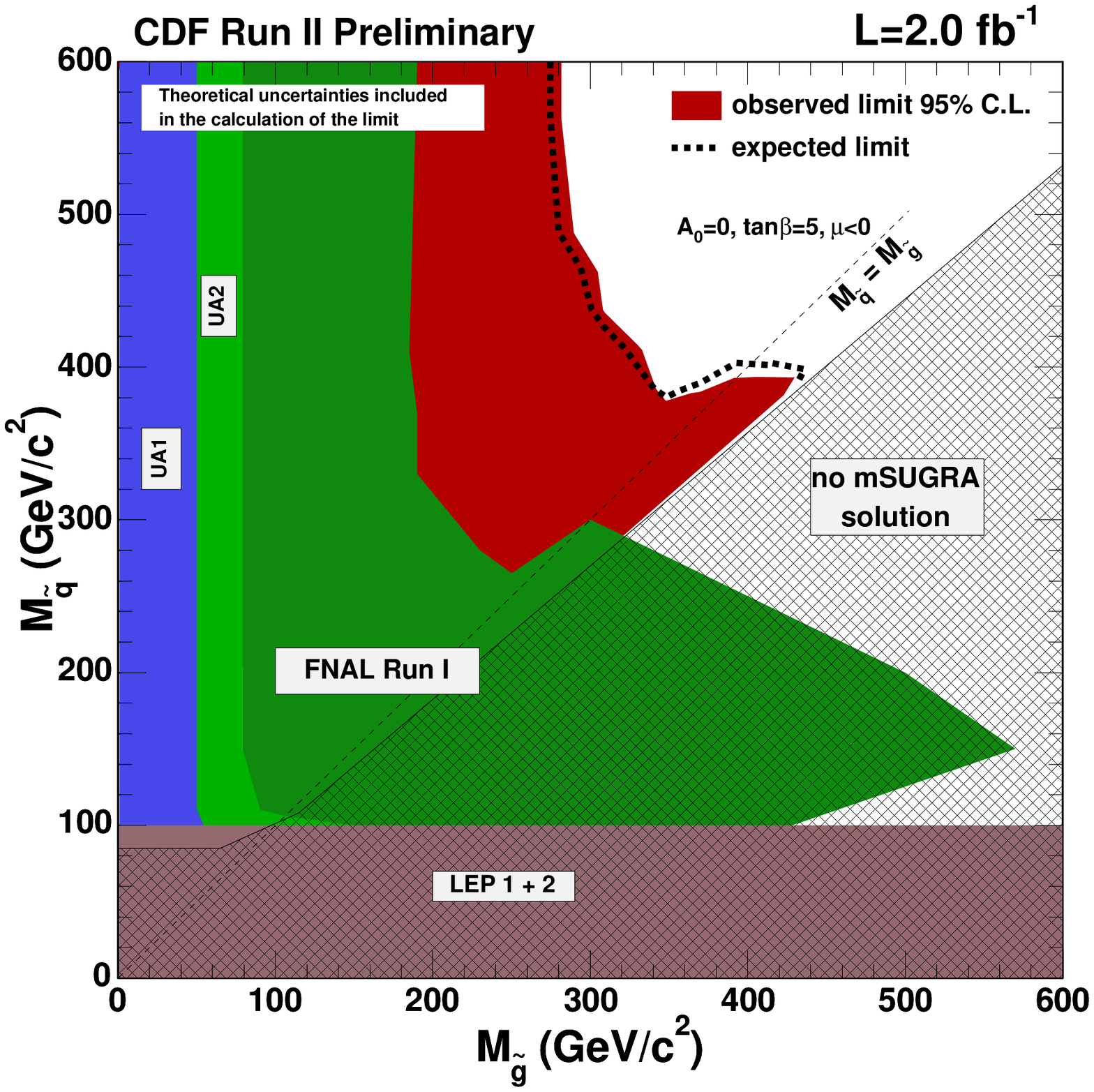}
\caption{The mSUGRA squark vs. gluino limit. \label{fig3}}
\end{minipage}\hspace{1pc}%
\begin{minipage}{18pc}
\includegraphics[scale=0.37]{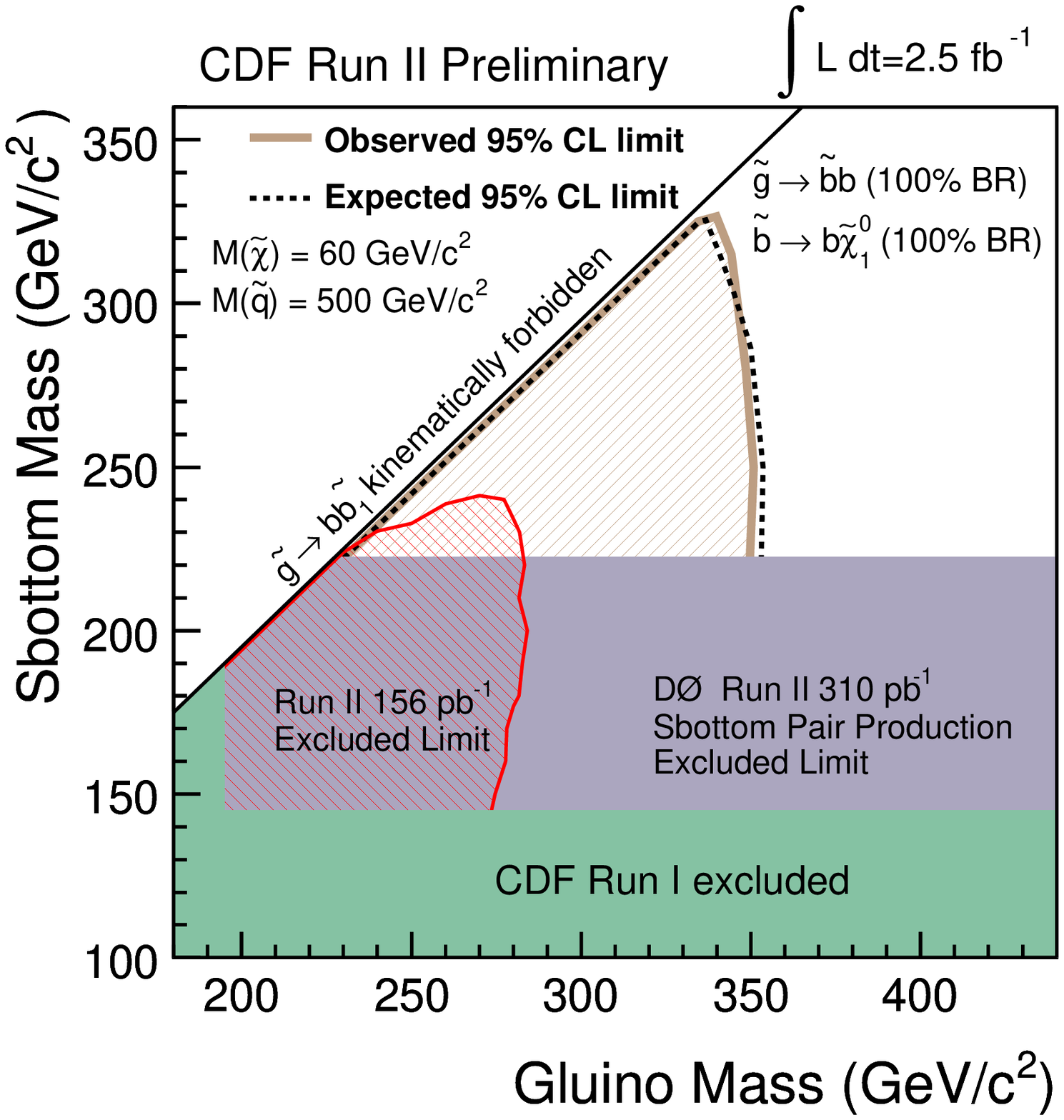}
\caption{The mSUGRA sbottom vs. gluino limit. \label{fig4}}
\end{minipage}
\label{wcharge}
\end{figure}

We analyze 2 fb$^{-1}$ of CDF data [2], requiring lepton veto (for top and boson reduction), small jet-met angle (for QCD reduction), high-$E_T$ jets, and high $H_T$ and $\met$.  We expect a background of $16 \pm 5$, $37 \pm 12$, $48 \pm 17$ dijet, trijet and 4-jet evens respectively, and we observe 18, 38 and 45 events, consistent with the SM prediction.  Figure \ref{fig3} shows the area in the squark-mass vs. gluino-mass parameter space that is excluded from this analysis at 95\% CL, along with previously set limits.

A special case of gluino pair production, is the one that results in two sbottom-bottom pairs.  The final signature is 4 $b$-jets and $\met$ from the LSPs.  By tagging the $b$-jets, we can significantly reduce the QCD background.  The respective 2.5 fb$^{-1}$ [3] CDF analysis, assumes 100\% BR of gluinos to sbottom quarks, which can be expected if the sbottom quarks are significantly lighter than the rest.  Two channels are investigated: 1 $b$-jet or $\ge$ 2 $b$-jets.  The main background is QCD (light- and heavy-flavor), with lower contributions from top pair production (significant in the 2-$b$-jets case) and ($W/Z$)+jets.  For the most sensitive 2-$b$-jets analysis we expect $2.3 \pm 0.8$ events and observe 2, which is interpreted as $M_{\tilde{g}}>350$ GeV/$c^2$ at 95\% CL, in the mSUGRA scenario assuming $M_{\tilde{q}}=500$ GeV/$c^2$ and $M_{\tilde{\chi}}=60$ GeV/$c^2$. Figure \ref{fig4} shows the excluded region in the $M_{\tilde{b}}$ vs. $M_{\tilde{g}}$ under the same assumptions.
\subsection{GMSB searches}
In the GMSB scenario the gaugino pair production will result in two photons, $\met$ and a number of leptons and/or jets.  The CDF analysis of 2.6 fb$^{-1}$ [4] requires two photons of energy above 13 GeV, separated by at least $\pi-0.15$ rads in the azimuthal, $\met$ significance above 3 and $H_T >200$ GeV, where $H_T$ now includes the $\met$.  The main SM backgrounds are electroweak with real $\met$ (63\%), QCD with fake $\met$ (37\%) and negligible non-collision background (beam halo and cosmics).  We expect $1.2 \pm 0.4$ SM events and we observe none.  Figure \ref{fig5} shows the current exclusion in the lightest neutralino lifetime vs. mass, as well as future projections with higher-luminosity, for several photon- and jet-multiplicities.

\subsection{ RPV searches}

If R-parity is violated, SUSY processes can result in lepton and baryon 
number violation.  At CDF, we investigate the $\tilde{\nu}$ lepton-number-violating decays to $e\mu$
or $e\tau$ or $\mu\tau$, using 1 fb$^{-1}$ of data with high-$p_T$ leptons
($>20$ GeV/$c$).  The main SM backgrounds are $Z\rightarrow\tau\tau$ (50\%), ($W/Z$)+fakes, diboson, top pair production and QCD ($\gamma$+jets), the latter being significant (20\%) in the $e\tau$ channel.  A control region between 50-110 
GeV/$c^2$ in the dilepton mass is used for validating these background estimations.  The 
dilepton mass cut that defines the signal region depends on the subanalysis. We predict
$0.1 \pm 0.1$, $1.4 \pm 0.3$, $1.0 \pm 0.3$ SM events in the $e\mu$
or $e\tau$ or $\mu\tau$ channels, and we observe 0, 2 and 2 respectively, as expected.
%\vspace{-0.1cm}
\section{Extra Dimensions}

\begin{figure}[t]
\begin{minipage}{18pc}
\includegraphics[scale=0.3]{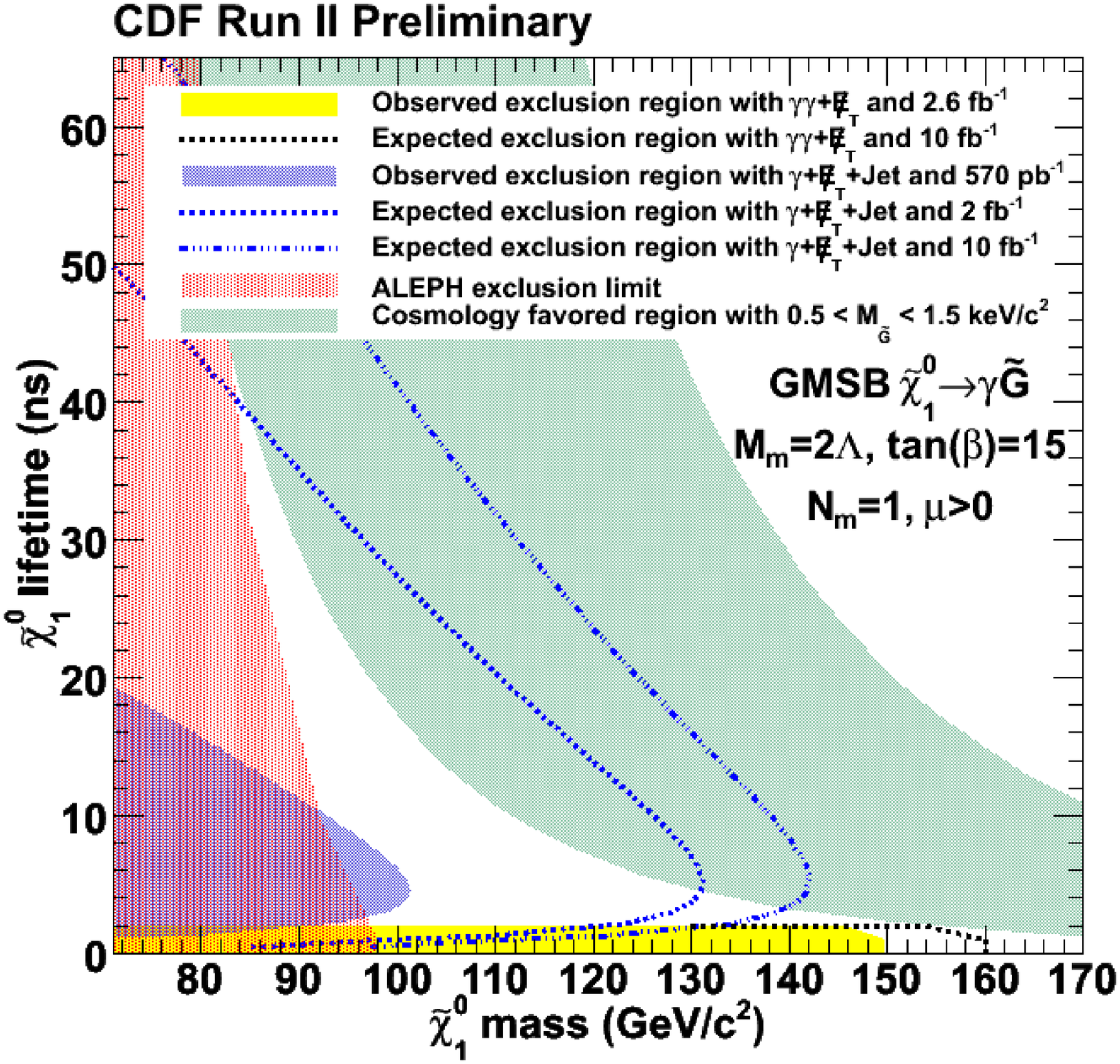}
\caption{GMSB neutralino mass vs. timelife limit.\label{fig5}}
\end{minipage}\hspace{1pc}%
\begin{minipage}{18pc}
\includegraphics[scale=0.45]{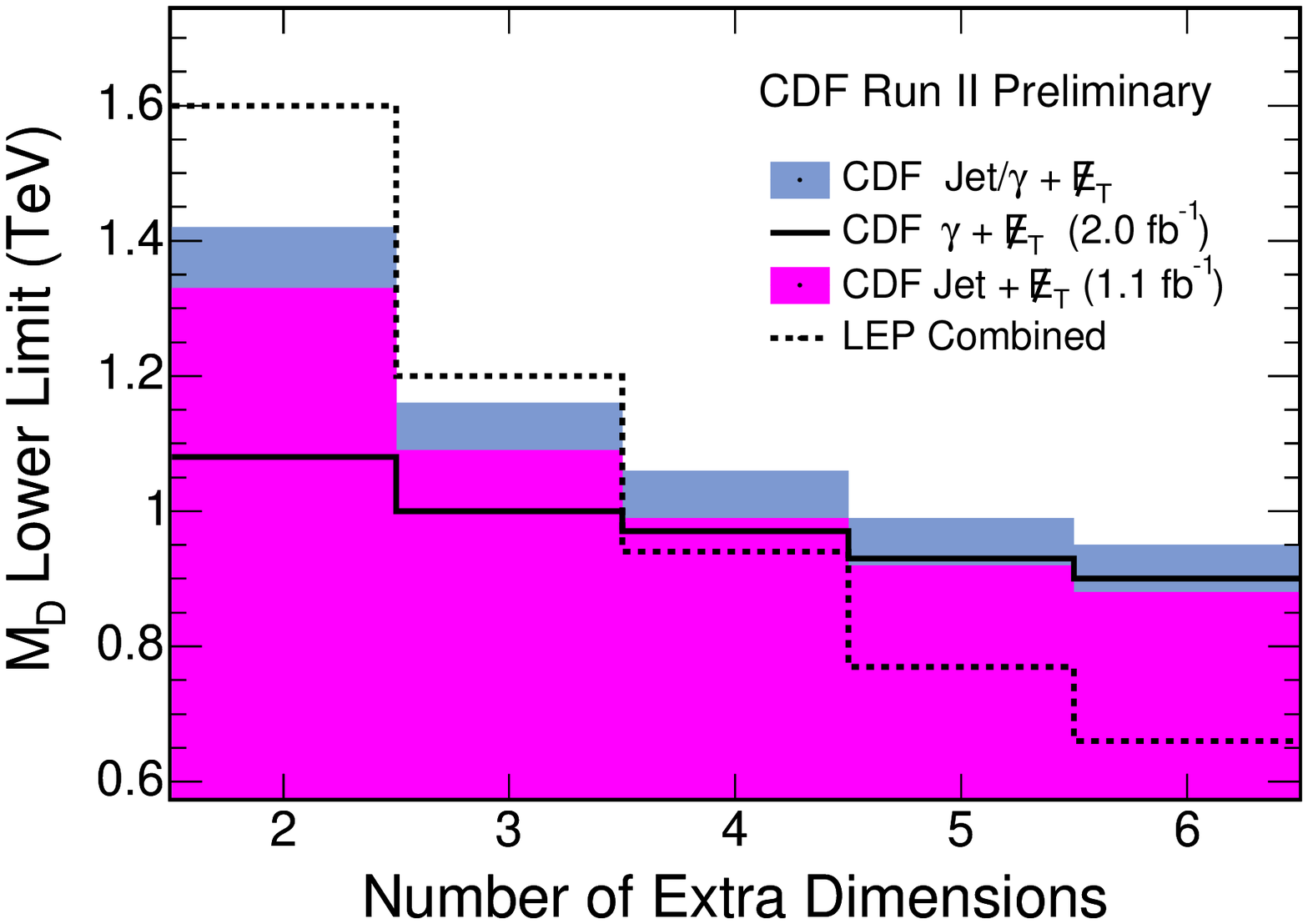}
\caption{Extra dimension mass scale vs. number of extra dimensions limit.\label{fig6}}
\end{minipage}
\label{wcharge}
\end{figure}

The hierarchy problem could be solved if gravity is allowed to move in extra dimensions.  If the graviton is produced in association with a photon or a parton, the resulting signature would be $\gamma+\met$ or jet+$\met$.
A recent 2 fb$^{-1}$ analysis looks for a photon with energy above 90 GeV and $\met$ 
above 50 GeV with no jets(tracks) above 15(10) GeV(GeV/$c$).  The 
main SM backgrounds are $(Z\rightarrow \nu\nu) +\gamma$ (56\%), cosmics (21\%), $W+\gamma$ (10\%), $W$+fake-$\gamma$ (8\%) and SM diphoton (5\%).  We expect $46 \pm 3$ SM events and we observe 40.  Combining with a previous CDF jet+$\met$ analysis [5], we set limits in the mass scale of large extra dimension vs. the number of extra dimensions, as shown in Figure \ref{fig6}.
%\vspace{-0.1cm}
\section{Heavy New Bosons}

The search for high-mass dilepton resonances is sensitive to new bosons of spin-0 (RPV sneutrino), spin-1 (GUT $Z'$), and spin-2 (RS graviton).  We have analyzed 2.3-2.5 fb$^{-1}$ of CDF dimuons [6] (dielectrons [7]) of opposite charge and $p_T$ above 30 (25) GeV/$c^2$.  The main backgrounds are DY, dibosons, $t\bar{t}$, and fakes.  The dimuon observation is consistent with background.  The dielectron spectrum demonstrates an excess at $\sim$240 GeV/$c^2$, which locally is a 3.8 $\sigma$ effect, whereas the probability to see it from 150 GeV to 1 TeV is 0.6\% (2.5 $\sigma$).  Our lower limits for the masses of $\tilde{\nu}$, $Z'$, $G_{\rm RS}$ are 1030 (963), 921 (850), 866 (397) GeV/$c^2$ for our $\mu\mu$ (ee) analyses, for specific model assumptions.
The spin-2 graviton could also decay to $ZZ$.  We investigate the $llll$ and $ll$+dijet final states (where $l$ is an electron or a muon) using 2.5-2.9 fb$^{-1}$ of data [8].  The main SM backgrounds are $W$+jets, $Z$+jets, QCD, and diboson.  Observation is consistent with expectation, and the graviton lower mass limit is 491 GeV/$c^2$ at 95\% CL for $k/M_p=0.1$.
%\begin{thebibliography}{0}
%
%\vspace{0.3cm}
\\
\begin{tabbing}
\=[1] http://www-cdf.fnal.gov/physics/exotic/ \hspace{0.6 cm} \= [5] CDF, Phys.\ Rev.\ Lett.\ {\bf 101}, 181602 (2009).\\
\>[2] CDF, Phys.\ Rev.\ Lett.\ {\bf 102}, 121801 (2009).\> [6] CDF, Phys.\ Rev.\ Lett.\ {\bf 102}, 091805 (2009).\\
\>[3] CDF, Phys.\ Rev.\ Lett.\ {\bf 102}, 221801 (2009). \> [7] CDF, Phys.\ Rev.\ Lett.\ {\bf 102}, 031801 (2009).  \\
\>[4] CDF, public note \#9625 (2009). \> [8] CDF, public note \#9640 (2009). 
\end{tabbing}

%\end{thebibliography}
\end{document}